\documentclass[pdftex,twocolumn,epjc3]{svjour3}
\RequirePackage[T1]{fontenc}
\usepackage{amsmath}
\RequirePackage{graphicx}
\RequirePackage{mathptmx}      
\RequirePackage{flushend}
\RequirePackage[numbers,sort&compress]{natbib}
\RequirePackage[colorlinks,citecolor=blue,urlcolor=blue,linkcolor=blue]{hyperref}
\usepackage{xcolor}
\usepackage{color}
\definecolor{darkgreen}{rgb}{0,0.5,0}
\definecolor{darkblue}{rgb}{0,0,0.7}
\definecolor{darkred}{rgb}{0.5,0,0.0}
\definecolor{darkorange}{rgb}{0.8,0.4,0.0}
\newcommand{\MSbar}{\overline{\text{MS}}}

\usepackage{amsmath}
\numberwithin{equation}{section}
\usepackage{amssymb}
\usepackage{xspace}
\usepackage{float}
\usepackage{subfig}
\usepackage{comment}
\journalname{Eur. Phys. J. C}
\DeclareMathOperator{\De}{d}
\newcommand{\de}{\De\!}
\newcommand{\as}{\alpha_{\text{S}}}

\newcommand{\Exp}{\text{Exp}}

\newcommand{\NLLp}{\text{NLL}^\prime}

\newcommand{\cf}{C_{\text F}}
\newcommand{\ca}{C_{\text A}}

\newcommand{\kt}{\vec{k}_t}
\newcommand{\gs}{\gamma_\text{soft}}
\newcommand{\gszero}{\gamma_\text{soft}^{(0)}}

\newcommand{\Sep}{S_\epsilon}

\newcommand{\muf}{\mu_{\text{F}}}

\newcommand{\muOf}{\mu_{0\text{F}}}
\newcommand{\mur}{\mu_{\text{R}}}

\newcommand{\muOr}{\mu_{0\text{R}}}

\newcommand{\calC}{\mathcal{C}}
\newcommand{\calD}{\mathcal{D}}
\newcommand{\ord}[1]{{\cal O}\left(#1\right)}

\usepackage{cprotect}

\begin{document}

\title{Bridging massive and massless schemes for soft gluon resummation in heavy-flavour production in $e^+e^-$ collisions}

\author{
Andrea Ghira\thanksref{e1,addr1}
\and
Lorenzo Mai\thanksref{e2,addr1}
\and
Simone Marzani\thanksref{e3,addr1}
}

\thankstext{e1}{e-mail: andrea.ghira@ge.infn.it}
\thankstext{e2}{e-mail: lorenzo.mai@ge.infn.it}
\thankstext{e3}{e-mail: simone.marzani@ge.infn.it}

\institute{
\label{addr1}
Dipartimento di Fisica, Universit\`a di Genova and INFN, Sezione di Genova, Via Dodecaneso 33, 16146, Italy
}

\date{}

\maketitle

\begin{abstract}

Perturbative calculations for processes involving heavy flavours can be carried out using two approaches: the massive and the massless schemes. 
These schemes can also be combined to leverage their respective strengths. Additionally, both massive and massless frameworks can be supplemented by soft-gluon resummation. However, matching resummed calculations across the two schemes presents significant challenges, primarily due to the non-commutativity of the soft and small mass limits.
The consistent resummation of mass and soft logarithms has been recently achieved at next-to-leading logarithmic (NLL) accuracy. 
In this paper, we consider heavy-quark fragmentation functions in electron-positron collisions and we extend this framework to achieve the so-called NLL$^\prime$ accuracy, which accounts for finite terms in the soft limit.

\end{abstract}

\section{Introduction}

Heavy flavours play a key part in particle physics phenomenology because of their role in various fundamental processes.  
In particular, not only the quarks beauty ($b$) and charm ($c$) 
are important for studies of the Higgs boson, but they also provide us with a powerful link between perturbative and non-perturbative Quantum Chromodynamics (QCD). 

From an experimental perspective, the long lifetimes of $B$ and $D$ hadrons ensure that their decays occur at measurable distances from the interaction point. This property is exploited by dedicated $b-$ and $c-$tagging techniques, which are widely used in collider experiments to identify $B$ and $D$ hadrons or $b-$ and $c-$jets.
From a theoretical perspective, calculations with identified heavy flavours are feasible primarily because the quark mass establishes a perturbative scale for the running coupling and simultaneously regulates collinear singularities.  The latter gives rise to the so-called dead-cone effect~\cite{Dokshitzer:1991fd,Dokshitzer:1995ev}, that is to say, a depletion of QCD radiation around the heavy-quark direction. 

Two approaches are commonly used for QCD calculations involving heavy flavours. In the massive scheme, heavy quarks are treated as on-shell massive particles, retaining full mass-dependent kinematics. This method has enabled calculations up to NNLO precision \cite{Czakon:2013goa,Catani:2019iny,Catani:2020kkl,Catani:2022mfv,Buonocore:2022pqq, Buonocore:2023ljm}, further combined with soft-gluon, high-energy, and transverse momentum resummations \cite{Cacciari:2011hy,Gaggero:2022hmv,Catani:1990eg, Ball:2001pq, Catani:2014qha}. By contrast, in the massless scheme, the mass is retained only as the regulator of collinear divergences and powers of $m^2/q^2$, with $m$ the mass of the heavy quark and $q$ the hard scale of the process, are systematically neglected. Within this framework, the differential cross-section is factorized in terms of massless partonic coefficient functions convoluted with universal fragmentation functions. The latter obey DGLAP evolution equations that allow to resum large logarithms of $m^2/q^2$. Unlike the case of light quarks, the initial conditions of the heavy-quark fragmentation functions are perturbatively calculable. By construction, they are free of mass logarithms but are affected by soft logarithms that should be resummed.

Since soft-gluon resummation is available in both the massive and massless schemes, it is natural to consider the possibility of matching these calculations at the resummed level to obtain a theoretical prediction that fully captures both mass and soft logarithms. However, as noted in the literature \cite{Gaggero:2022hmv, Aglietti:2006wh, Aglietti:2022rcm, Cacciari:2002re, Corcella:2003ib}, this task is far from straightforward. The structure of soft logarithms differs in the two approaches, posing substantial challenges to the development of a consistent all-order matching framework. The formalism that allows for the consistent resummation of both mass logarithms and soft logarithms of the Mellin variable $N$ was developed in \cite{Ghira:2023bxr} at NLL accuracy (see also~\cite{Aglietti:2006wh, Aglietti:2022rcm}).
In this paper, we extend the formalism to include the $\ord{\as}$ contribution that does not vanish in the large-$N$ limit, thus reaching the so-called NLL$^\prime$ accuracy. Although the framework we develop is rather general, in this paper we consider fragmentation functions in $e^+ e^-$ collisions. Furthermore, this study focuses only on parton-level results, leaving detailed phenomenological analyses for future work.

The paper is organized as follows. In Sec.~\ref{sec: hf overview} we review the theoretical framework
of soft-gluon resummation for heavy quark production in $e^+ e^-$ annihilation for both massive and massless schemes. We also review the non-commutativity of the soft and massless limits and how this can be overcome for the logarithmic structure.  
The calculation that allows us to go beyond NLL accuracy is performed, in momentum space, in Sec.~\ref{sec: our computation}, while its inclusion in the Mellin-space resummed expression is described in Sec.~\ref{sec: numerics}, together with a study of its numerical impact. Finally, we draw our conclusions and future prospects in Sec.~\ref{sec: conclusion}. Details of the calculations and explicit results are collected in the Appendices.

\section{Heavy quark production in $e^+e^-$ collisions}\label{sec: hf overview}
In this section, we analyse the production of a heavy quark-antiquark pair of momenta $p_1,p_2$  plus undetected radiation $X$, in electron-positron collisions:
\begin{equation}
    e^+ e^- \to \gamma/Z \to h(p_1)+\bar{h}(p_2)+X(k).
\end{equation}
We consider the differential cross section $\frac{\de \sigma}{\de x}$, with $x= \frac{2 p_1 \cdot q}{q^2}$ and $q$ is the $e^+e^-$ centre-of-mass energy.
The calculation can be carried out in two different schemes, namely the massive scheme and the massless one.

In the massive approach, the spectrum $\frac{\de \sigma}{\de x}$ is computed to a fixed order in perturbation theory, treating all dependence on the mass $m$ of the heavy quark exactly. In this framework, collinear singularities are regulated by the heavy quark mass, leading to the appearance of logarithmic terms $\log \frac{m^2}{q^2}$ in the perturbative coefficients. These logarithms can grow large when $q^2 \gg m^2$, potentially undermining the behaviour of the perturbative expansion. Nonetheless, this approach allows for an exact treatment of radiation kinematics across the full phase space, up to the specified order in $\as$.

In the massless scheme, the mass of the heavy quark is assumed to be much smaller than the hard scale of the process, i.e.\ $m^2 \ll q^2$. In this framework, heavy-quark production is described through the collinear factorization formula:
    \begin{equation}
    \label{eq: factorisation formula}
    \frac{1}{\sigma_0}\frac{\de \sigma}{\de x}= \sum_i \mathcal{C}_i\left(x,\frac{\muf^2}{q^2},\frac{\mur^2}{q^2},\as(\mur^2)\right)\otimes \mathcal{D}_{i\to h}\left(x,\muf^2,m^2\right),
    \end{equation}
    with $\sigma_0$ the Born-level massless cross section and $\muf, \mur$ the factorization and renormalization scales, respectively. The sum in Eq. (\ref{eq: factorisation formula}) runs over all possible partonic channels.
The functions $\mathcal{C}_i$ represent process-dependent partonic cross sections, which can be expanded perturbatively in powers of $\as$ provided that $q^2$ is sufficiently large. These functions are convoluted with the process-independent fragmentation functions $\mathcal{D}_{i \to h}$.  
Collinear logarithms of $m^2/q^2$ are resummed to all orders, achieving a specified level of logarithmic accuracy via the solution of the DGLAP equations for the fragmentation functions.  
Additionally, the initial condition for the evolution equation of the heavy-quark fragmentation function is set at a scale of the order of the heavy quark mass, enabling a perturbative computation~\cite{Mele:1990cw, Melnikov:2004bm,Cacciari:2001cw}.

The convolution product in Eq.~(\ref{eq: factorisation formula}) simplifies to an ordinary product by applying the Mellin transformation 
\begin{equation}\label{eq: Mellin trasnform}
\tilde{f}(N) = \int_0^1 dx \, x^{N-1} f(x).
\end{equation} 
We then obtain
\begin{align}
\tilde{\sigma}(N, \xi) &= \frac{1}{\sigma_0} \int_0^1 \de x \, x^{N-1} \frac{\de\sigma}{\de x} = \\\nonumber
&\sum_i \tilde{\calC}_i \left( N, \frac{\muf^2}{q^2}, \frac{\mur^2}{q^2}, \as(\mur^2) \right) \tilde{\calD}_i(N, \muf^2, m^2), 
\end{align}
where the scale dependence of the fragmentation functions is governed by the DGLAP evolution equations
\begin{equation}\label{eq: DGLAP equation}
\muf^2 \frac{\de}{\de\muf^2} \tilde{\calD}_i(N, \muf^2, m^2) = \sum_j \gamma_{ij}\left( N, \as(\muf^2) \right) \tilde{\calD}_j(N, \muf^2, m^2),
\end{equation}
where the anomalous dimensions $\gamma_{ij}$, which are the Mellin transforms of the time-like splitting functions, can be computed perturbatively. The solution of Eq.~(\ref{eq: DGLAP equation}) can be expressed schematically in terms of an evolution kernel matrix $\tilde{E}_{ij}$ and initial conditions defined at a reference scale $\muOf$: 
\begin{equation}
\tilde{\calD}_i(N, \muf^2, m^2) = \sum_j \tilde{E}_{ij}(N, \muOf^2, \muf^2) \tilde{\calD}_{0j}(N, \muOf^2, m^2).
\end{equation}

In the following, we denote by $\tilde{\sigma}^{(n_f)}_k(N, \xi)$  the Mellin transformed cross section computed in the massless scheme, and by $\tilde{\sigma}^{(n_f-1)}_k(N, \xi)$ the corresponding cross-section computed in the massive scheme, with $\xi =m^2/q^2$.  
Here, $n_f$ denotes the number of active flavours at the hard scale $q^2$. 
On the one hand, both the coefficient functions $\tilde{\calC}_i$ and the initial conditions for the fragmentation functions $\tilde{\calD}_{0j}$ in $\tilde{\sigma}^{(n_f)}_k(N, \xi)$ are computed at order $\as^k$. The evolution equations for the fragmentation functions  are solved at (next-to)$^\ell$-leading logarithmic (N$^\ell$LL) accuracy and one typically chooses $\ell=k$. 
On the other hand, the cross section $\tilde{\sigma}^{(n_f-1)}_k(N, \xi)$ is computed at order $\as^k$ with full mass dependence. 
By combining these two schemes, one can take advantage of their respective strengths:
\begin{equation}
\label{eq: FONLL}
\tilde{\sigma}_k(N, \xi) = \tilde{\sigma}^{(n_f-1)}_k(N, \xi) + \tilde{\sigma}^{(n_f)}_{k}(N, \xi) - \text{double counting},
\end{equation}
where "double counting" refers to the perturbative expansion of $\tilde{\sigma}^{(n_f)}_k(N, \xi)$ up to order $k$. In this work, we restrict our focus to the case $k = 1$, known as the FONLL scheme \cite{Cacciari:1998it}.
\subsection{Large $N$ behaviour}

Both quantities appearing on the r.h.s. of Eq. (\ref{eq: FONLL}) display
a logarithmically enhanced behaviour at large $N$, which corresponds to the limit $x\to 1$ in the physical space. These contributions can be resummed to all order in QCD up to a certain logarithmic accuracy.
As already noted in \cite{Gaggero:2022hmv, Ghira:2023bxr}, the logarithmic structure of $\tilde{\sigma}^{(n_f-1)}$ and $\tilde{\sigma}^{(n_f)}$ differ in the large $N$ limit. Specifically, the order-$\as^n$ perturbative coefficient in the expansion of $\tilde{\sigma}_k^{(n_f-1)}$, with $n\leq k$, is a polynomial of degree $n$ in $\log N$, plus terms that vanish as $N \gg 1$. 
The coefficients of this polynomial retain the full $\xi$ dependence: they include both powers of $\xi$, which vanish in the massless limit, and powers of $\log \xi$. 

The coefficient functions $\tilde{\calC}_i$ in $\tilde{\sigma}_k^{(n_f)}$ are ordinary, subtracted partonic cross sections in massless QCD. The perturbative coefficients here contain up to two powers of $\log N$ for each power of $\as$, which corresponds to residual effects of soft singularity cancellation and collinear singularity subtraction.
Finally, the initial conditions for the evolution of the fragmentation functions, $\tilde{\calD}_{0j}$, can be computed perturbatively. The associated order-$\as^n$ coefficients are polynomials of degree $2n$ in $\log N$, with coefficients that depend on $\log \xi$ but are independent of powers of $\xi$.

As mentioned in the introduction, the resummation of soft-gluon contributions to all orders, up to a given logarithmic accuracy, for both $\tilde{\sigma}_k^{(n_f-1)}(N, \xi)$ and $\tilde{\sigma}_k^{(n_f)}(N, \xi)$ can be performed and has been studied in the past.  
The combination of these two formalisms was carried out in \cite{Ghira:2023bxr} at NLL accuracy in both $\log \xi$ and $\log N$. In the following, we concisely review this setup. 
Let us start by considering soft-gluon resummation in the massless scheme. Within this picture, the resummation of soft logarithms for the $b$-quark energy spectrum was first performed in Ref.~\cite{Cacciari:2001cw}, with the specific case of $H \to b \bar{b}$ studied in Ref.~\cite{Corcella:2004xv}.  
In the large $N$ limit, we restrict ourselves to the non-singlet channel of the fragmentation functions as it is the only source of enhanced $\log N$ corrections. Within this approximation, the resummed expression for the energy spectrum at N$^k$LL accuracy in the logs of the mass and at N$^{\ell_1}$LL accuracy in the logs of $N$, takes the general form:
\begin{align}\label{fragm-master}
		&\widetilde{\sigma}_{k \ell_1}^{(n_f,\,\text{res})}(N,\xi)= \widetilde{\mathcal{E}}(N,\muOf^2,\muf^2) 
        \nonumber \\
      &\left(1+ \sum_{n=1}^{\ell_1}\left(\frac{\as(\mur^2)}{\pi}\right)^n\mathcal{C}_0^{(n)} \right) \left(1+ \sum^{\ell_1}_{n=1}\left(\frac{\as(\muOr^2)}{\pi}\right)\mathcal{D}_0^{(n)} \right) 
                   \nonumber\\ &
                   \Exp\left[{C\left(N,\frac{\muf^2}{q^2},\frac{\mur^2}{q^2},\as(\mur^2)\right)}\right] \nonumber \\
                   & \Exp\left[{D_{0}\left(N,\frac{\muOf^2}{m^2},\frac{\muOr^2}{m^2},\as(\muOr^2)\right)}\right].
	\end{align}
Here, $\Exp\left[{C}\right] $ and $\Exp \left[D_{0}\right]$ denote the resummed logarithmically enhanced part of the coefficient function and of the initial condition respectively, while $\widetilde{\mathcal{E}}$ represents the DGLAP evolution kernel for the fragmentation function. The renormalization and factorization scales $\mur^2,\muf^2$ are chosen to be of the order of magnitude of $q^2$, while the corresponding reference values
 $\muOr^2,\muOf^2$ are of order  $m^2$.
The coefficients $\mathcal{C}_0^{(n)}$ are process dependent while $\mathcal{D}_0^{(n)}$ are process independent. Their inclusion up to the $\ell_1$-th order allows one to achieve the N$^{\ell_1}$LL$^\prime$ accuracy.
For the case of interest, at the lowest in perturbation theory they read \cite{Altarelli:1979kv, Cacciari:2001cw}:
\begin{align}\label{eq: constants massless}
    \mathcal{C}_0^{(1)} &= \cf\left(\frac{5}{12}\pi^2 -\frac{9}{4} + \frac{3}{4} \log \frac{q^2}{\muf^2}\right), \nonumber \\
    \mathcal{D}_0^{(1)} &= \cf\left(-\frac{\pi^2}{6} +1+ \frac{3}{4} \log \frac{\muOf^2}{m^2}\right).
\end{align}

The general formalism to perform soft-gluon resummation in the massive scheme was described in~\cite{Laenen:1998qw}. The all-order N$^{\ell_2}$LL expression takes the exponentiated form 
\begin{align}\label{eq:massive resummation}
	&\widetilde{\sigma}^{(n_f-1,\,\text{res})}_{k\;\ell_2}(N,\xi)=\left(1+K(\xi,\as)\right) \nonumber \\
    &\Exp\left[{2\int^1_0 \de x \frac{x^{N-1}-1}{1-x} \gs \left(\xi,\as\left((1-x)^2 q^2\right)\right)}\right].
\end{align}
The process-dependent factor $K(\xi,\as)$ is free of soft logarithms and is computed up to order $\as^k$. 
At $\ord{\as}$ (for the process of interest) it reads\footnote{At $\ord{\as}$ and in the small mass limit the result is independent on the exchanged vector boson.} 
\begin{align} 
\label{eq: K as}
&K(\xi,\as(\mur^2)) = \frac{\as(\mur^2)\cf}{\pi} \nonumber \\
&\left(\frac{1}{2}\log^2 \xi-\frac{1}{2}\log \xi +\frac{\pi^2}{2}-1+\ord{\xi}\right)+\ord{\as^2}.
\end{align}
The massive soft anomalous dimension is instead process-independent
\begin{align}
\label{eq:gs0}
\gs(\xi,
\as(\mur^2)) =\frac{\as(\mur^2)}{\pi}\gszero(\beta)+\mathcal{O}(\as^2),
\end{align}
with
\begin{equation}
\beta=\sqrt{1-4\xi},
\end{equation}
and 
\begin{equation}
	\gszero(\beta)=\cf\left(\frac{1+\beta^2}{2\beta}\log\frac{1+\beta}{1-\beta}-1\right).
	\label{eq:gammazero}
\end{equation}

Already at this stage one can verify that the logarithms arising in Eq. \eqref{fragm-master} and those from the massless limit of Eq. \eqref{eq:massive resummation} are different; the same holds also for the constants in Eq. \eqref{eq: constants massless} and Eq. \eqref{eq:gs0}. 
Focusing on the former, we note that the resummed expression in Eq.~(\ref{eq:massive resummation}) features, at most, single logarithms of $N$ to any order in perturbation theory, i.e.\ $\as^n \log^m N$, with $m\le n$. This is not surprising: collinear singularities are absent because of the finite quark mass and, consequently, collinear logarithms do not appear as logarithms of $N$ but rather as logarithms of the mass.

The merging between Eq.~\eqref{fragm-master} and Eq.~\eqref{eq:massive resummation} into a consistent NLL resummation formula can be performed by first noting that the massless result Eq.~\eqref{fragm-master} can be recast in terms of two jet functions $J, \bar J$:
\begin{align}\label{eq:FF-sep}
		&\widetilde{\sigma}_{k=1, \ell_1=1}^{(n_f,\,\text{res})}(N,\xi)=  \widetilde{\mathcal{E}}^{(\text{sub})}(N,\muOf^2,\muf^2) 
         \nonumber\\ &
				 \left(1+ \frac{\as(\mur^2) }{\pi}\mathcal{C}_0^{(1)} \right) \left(1+ \frac{\as(\muOr^2) }{\pi}\mathcal{D}_0^{(1)} \right) 
                   \nonumber\\ &
				 \Exp \left[J\left(N,q^2,\mur^2,\muf^2,\muOr^2,\muOf^2\right)+\bar{J}\left(N,q^2,\mur^2\right)\ \right].
\end{align}
The function \( J \) describes the dynamics associated with the collinear radiation off the measured quark $h$, while \( \bar{J}\) characterizes the physics related to the unmeasured (recoiling) particle $\bar h$. In Eq.~\eqref{eq:FF-sep}, $\Exp \left[J\right]$ is related to the product of the initial condition, the $\log N$ enhanced part of the evolution operator and a term of the coefficient functions, while $\Exp \left[\bar J\right]$ is related to the remaining part of the coefficient function. Lastly, the factor $\widetilde{\mathcal{E}}^{(\text{sub})}$ accounts for the contribution from DGLAP evolution that is not enhanced at large $N$. 
Explicit expressions are collected in \ref{app: nll explicit formulas}.

The construction of resummation scheme that consistently accounts for both soft and mass logarithms stems from the observation that while $J$ is computed in the quasi-collinear limit~\cite{Catani:2000ef,Catani:2002hc}, i.e.\ the mass of the heavy quark and the transverse momentum of the gluon radiation are both small but of the same order, $\bar J$ is calculated in the massless approximation. 
The main result of Ref.~\cite{Ghira:2023bxr} is to relate the jet functions $J$ and $\bar J$ to the exponentiation of the cumulative distributions $j$ and $\bar j$, both computed in the quasi-collinear limit:
\begin{align}\label{eq:jet-function-j}
j(N,\xi)=& -\int_0^1 \de z_1 \int_{z_1^2 m^2}^{q^2} \frac{\de \kt^2}{\kt^2} \frac{\as^\text{CMW}(\kt^2)}{2\pi} \nonumber \\
\times&P_{gh} (z_1,\kt^2-z_1^2 m^2) \Theta\left(\eta_1\right)   \Theta\left(z_1 -\frac{1}{\bar N} \right),  \\ \label{eq:jet-function-jbar}
\bar j(N,\xi) =&-\int_0^1 \de z_2\int_{z_2^2 m^2}^{q^2} \frac{\de \kt^2}{\kt^2} \frac{\as^\text{CMW}(\kt^2)}{2\pi} \nonumber \\
&P_{gh} (z_2,\kt^2-z_2^2 m^2)\Theta\left( \eta_2\right)  \Theta\left( \frac{\kt^2}{q^2 z_2} -\frac{1}{\bar N}\right),
\end{align}
with $\bar N= N e^{\gamma_{\text{E}}}$ and
\begin{align}
    P_{gh}(z,\kt^2)= \cf\left(\frac{1+(1-z)^2}{z}-\frac{2 m^2 z (1-z)}{\kt^2+z^2 m^2}\right).
\end{align}
Here, CMW refers to the Catani-Marchesini-Webber scheme \cite{Catani:1990rr}:
\begin{equation} \label{eq:decoupling}
    \as^{\text{CMW}}(\kt^2)= \as(\kt^2)\left(1+\frac{\as(\kt^2) K^{(n)}}{2\pi}\right),
\end{equation}
where $n$ is the number of active flavours at the scale $\kt^2$ and $K^{(n)}= \ca \left(\frac{67}{18}-\zeta_2\right)-\frac{5}{9} n$. The expression for $\as(\kt^2)$ is given by
\begin{align}
    \as(\kt^2) &= \as^{(5)}(\kt^2)\Theta(\kt^2-m_b^2) \nonumber \\
    &+ \as^{(4)}(\kt^2)\Theta(\kt^2-m_c^2)\Theta(m_b^2-\kt^2) \nonumber \\
    &+ \as^{(3)}(\kt^2)\Theta(m_c^2-\kt^2),
\end{align}
with:
\begin{align}\label{eq:rc-nf}
    &\as^{(n)}(\kt^2)= \frac{\as^{(n)}(\mu_{n}^2)}{\ell}\left(1-\frac{\beta_1^{(n)}}{\beta_0^{(n)}}\frac{\log \ell}{\ell}\right), \nonumber \\
    &\ell= 1+\as^{(n)}(\mu_{n}^2)\beta_0^{(n)}\log{\frac{\kt^2}{\mu_{n}^2}}.
\end{align}
Here $\beta_0^{(n)}= \frac{11 \ca-2 n}{12 \pi}$ and $\beta_1^{(n)}= \frac{17 \ca^2- 5n \ca-3n \cf}{24 \pi^2}$ denote the one and two loop coefficient of the QCD $\beta$- function. 
In what follows, the factorization scales $\muf$ and $\muOf$ are set to $q$ and $m$ respectively. Furthermore, we will keep the dependence on renormalization scales $\mur\simeq q$ and $\muOr \simeq m$ explicit in our expressions, but we shall leave the additional dependence on the renormalization scales $\mu_{n}$, which appear in Eq. \eqref{eq:rc-nf}, i.e.\ related to the decoupling scheme for the running coupling, understood.

The integrals of the quasi-collinear splitting functions over the running coupling Eq.~(\ref{eq:decoupling}) give rise to resummed expressions that are defined piecewise. The different regions that one obtains depend on whether we are considering $b$ or $c$ fragmentation and on the hierarchy between the various scales, i.e.\ $\frac{1}{\bar N},\frac{1}{\bar {N}^2}, \frac{m_b^2}{q^2}$, and $\frac{m_c^2}{q^2}$, see e.g.~\cite{Ghira:2023bxr, Caletti:2023spr, Cacciari:2024kaa}.
The result smoothly interpolates between the soft-gluon resummed expression in the massless scheme, Eq.~\eqref{eq:FF-sep}, and the one in the massive scheme, Eq.~\eqref{eq:massive resummation}, at NLL in both soft and mass logarithms:~\footnote{Note that the evolution factor $\widetilde{\mathcal{E}}^{(\text{sub})}$ is beyond the scope of this framework since it does not contribute to the large-$N$ limit. However,  we have decided to keep it because it allows us to reproduce the full (non-singlet) DGLAP evolution.}
\begin{align}\label{eq:FF-sep modified NLL}
		&\widetilde{\sigma}^{(\text{NLL})}(N, \xi)
        =  \widetilde{\mathcal{E}}^{(\text{sub})}(N,m^2,q^2) 
                   \nonumber\\ &
				 \Exp \left[j\left(N,\xi,\frac{\mur^2}{q^2},\frac{\muOr^2}{m^2}\right)+\bar{j}\left(N,\xi,\frac{\mur^2}{q^2},\frac{\muOr^2}{m^2}\right)\ \right].
\end{align}
Note that in this resummed expression we have not included any finite $\ord{\as}$ contribution. Indeed, the constant contributions in $K$, Eq.~\eqref{eq: K as}, do not coincide with the sum of the constants in the massless scheme, Eq.~\eqref{eq: constants massless}. This discrepancy is not resolved by the NLL calculation of~\cite{Ghira:2023bxr} and prevents us from claiming NLL$^\prime$ accuracy.
In the next section, we address this problem by computing the $\ord{\as}$ corrections to heavy quark production.

\section{Momentum-space calculation} \label{sec: our computation}

To understand the physical origin of the different constant terms in Mellin space between the two formalisms, we find it convenient to go back to momentum space. Therefore, we perform the calculation of the heavy-quark energy spectrum at $\ord{\as}$, focusing on the contributions that are enhanced when $x \to 1$.  
Before describing this computation, it is useful to recall the limiting behaviour in the massless and massive schemes, at the same order.

First, we write expansion to $\ord \as$ of the inverse Mellin transform of Eq.~(\ref{fragm-master}):
    \begin{align}
    \label{eq: CC-fixed order}
    &\frac{1}{\sigma_0} \frac{\de \sigma^{(n_f, \, \text{f.o.})}}{\de x}= \delta(1-x)+\frac{\as \cf}{2\pi} 
    \nonumber \\
     &\left[ \delta(1-x)\left(\frac{2\pi^2}{3} -\frac{5}{2}\right) - \log \xi \left(\frac{1+x^2}{(1-x)_+}
     +\frac{3}{2}\delta(1-x)\right) \right.
     \nonumber \\
    & \left.-\frac{7}{2}\frac{1}{(1-x)_+}
    -2\left( \frac{\log (1-x)}{1-x} \right)_+ 
  +\dots\right],
\end{align}
where we recall that $\xi$ accounts for the heavy quark mass and the dots refer to contributions that are less-enhanced  in the $x\to 1$ limit, while the plus distributions are defined through their action on a test function $g(x)$ as
\begin{align}
    \int_0^1 \de x\, \left(\frac{\log^k(1-x)}{1-x}\right)_+ g(x) &\equiv \int _0^1 \de x  \,\frac{\log^k(1-x)}{1-x} \nonumber \\&\times[g(x)-g(1)].
\end{align}
Let us briefly comment on the expression in Eq.~\eqref{eq: CC-fixed order}. First, we recognize the regularized DGLAP splitting function as the coefficient of the mass logarithm. Additionally, we recall that the delta-function contributions correspond to constants in Mellin space, while the last $[\ldots]_+$ term in Eq. \eqref{eq: CC-fixed order} generates the double logarithms in Mellin space reviewed above. As already discussed in \cite{Ghira:2023bxr}, this term originates only from $\bar J$.

Second, we consider the $\ord \as$ contribution to the resummation in the massive scheme, Eq~\eqref{eq:massive resummation}. Going back to $x$-space and taking the $\xi \to 0$ limit, we find:
\begin{align}\label{eq: full fo}
     &\frac{1}{\sigma_0} \frac{\de \sigma^{(n_f-1, \, \text{f.o.})}}{\de x}= \delta(1-x)+\frac{\as \cf}{2\pi} \\
     &\left[\delta(1-x)\left(\log^2 \xi-\log \xi +\pi^2-2\right)-4\frac{ \log \xi+1}{(1-x)_+} + \dots\right],\nonumber 
\end{align}
where the dots now indicate contributions that are less enhanced in the $x \to 1$ limit or that are power-suppressed at small $\xi$. As already noted in~\cite{Gaggero:2022hmv}, the expression in Eq.~\eqref{eq: full fo} has a milder behaviour for $x \to 1$ than Eq.~\eqref{eq: CC-fixed order}, but features double logarithms of the mass. 

\subsection{Details of the computation}

In this section we compute the differential cross section $\frac{1}{\sigma_0}\frac{\de \sigma}{\de x}$ at $\ord{\as}$. 
We work in the $x\to 1$ limit, in which the calculation can be expressed as the sum of the virtual, soft, collinear and anti-collinear terms. The overlapping between the soft and collinear sectors will be removed at the end of the calculation.  
Specifically, we are interested in the small-mass limit of this calculation. This is straightforward for the virtual contribution and the soft sector, while it requires employing the quasi-collinear limit for the remaining collinear and anti-collinear regions. 
Furthermore, in order to keep track of the role played by the mass of the measured heavy quark and the mass of the recoiling one, we introduce $\xi_1= p_1^2/q^2=m_1^2/q^2$ and $\xi_2=p_2^2/q^2=m_2^2/q^2$.
~\footnote{Note that with different masses, we have that $x<1+\xi_1-\xi_2$. In what follows, in order to work with a scaling variable that ranges between zero and one, we implicitly perform the rescaling $x\to \frac{x}{1+\xi_1-\xi_2}$.}

\subsubsection{Virtual}

Working in $d=4-2\epsilon$ dimensions, the $\MSbar$ renormalized virtual contribution reads:
\begin{align}
    \frac{1}{\sigma^{(d)}_0}&\frac{\de \sigma^{(V)}}{\de x}= \frac{\as \cf S_\epsilon}{2\pi}~\delta(1-x)\left[- \frac{1}{\epsilon}\log\xi_1\xi_2  -\frac{2}{\epsilon} +\frac{4\pi^2}{3} -4 \right.
  \nonumber    \\
    &\left.-\frac{1}{2}\log\xi_1\xi_2\ + \frac{1}{2}\log^2\xi_1+ \frac{1}{2}\log^2\xi_2+\ord{\xi_1, \xi_2}\right] .
\end{align}
For convenience we have defined $\Sep= \left(\frac{4\pi \mur^2}{q^2}\right)^\epsilon\frac{1}{\Gamma(1-\epsilon)}$ and denoted with $\sigma_0^{(d)}$ the Born level massless cross section in $d$ dimensions. The result was checked against the small mass limit of \cite{Bonciani:2003ai, Bernreuther:2004ih} and with \cite{terHoeve:2023ehm}.

\subsubsection{Collinear and anti-collinear}

In this section, we derive the collinear and anti-collinear contributions. Specifically, we consider $1-x$ and $\xi_{1,2}$ small, but of the same order.

We begin by studying the QCD splitting process $h\to h(p)+g(k)$ in the quasi-collinear limit, being $h$ a generic heavy (anti)quark with $p^2=m^2$. The kinematics of the splitting can be formulated using the Sudakov parametrization:
\begin{align}\label{eq:sudakov_parametrisation_p1_k}
	p^\mu=& (1-z) P^\mu-k_t^\mu + \frac{\kt^2 + (1-(1-z)^2)m^2}{2(1-z) P\cdot n} n^\mu, \nonumber \\
	k^\mu=& z P^\mu+k_t^\mu + \frac{\kt^2 - z^2m^2}{2 z P\cdot n} n^\mu
\end{align}
where $P^\mu$ is a timelike four vector ($P^\mu P_\mu=m^2$) that denotes the forward direction while $n^\mu$ is a lightlike four vector that indicates the backward direction. Finally, the four vector $k_t^\mu$ is spacelike  with $k_t^\mu k_{t \mu}=-\kt^2$ and it is orthogonal to both $P^\mu$ and $n^\mu$. 
In the quasi-collinear limit, the following factorization holds
\begin{align}\label{eq:qc_fact}
    |\mathcal{M}|^2\simeq |\mathcal{M}_0|^2 \frac{8\pi \as z(1-z)}{\kt^2+z^2 m^2}P^{(d)}_{gh}(z,\kt^2),
\end{align} 
with $P^{(d)}_{gh}$ denoting the timelike massive splitting function in $d$ dimensions:
\begin{align}
    P^{(d)}_{gh}(z,\kt^2)= \cf\left(\frac{1+(1-z)^2}{z}-\frac{2 m^2 z (1-z)}{\kt^2+z^2 m^2}-\epsilon z\right).
\end{align}

We now exploit quasi-collinear factorization to compute the collinear and anti-collinear sectors. 
The collinear function is obtained by integrating Eq.~(\ref{eq:qc_fact}) against the gluon phase-space
\begin{align}
\de \Phi = \frac{1}{16 \pi^2} \frac{1}{\Gamma(1-\epsilon)}\left(\frac{4\pi \mur^2}{\kt^2}\right)^\epsilon \de \kt^2 \frac{\de z}{z (1-z)},
\end{align}
and the appropriate measurement function. We have
\begin{align}\label{eq:function-j}
\frac{1}{\sigma^{(d)}_0}\frac{\de \sigma^{(C)}}{\de x}&= \frac{\as  }{2\pi} \frac{(4\pi \mur^2)^\epsilon}{\Gamma(1-\epsilon)} \int_0^1 \de z_1\int_{0}^{q^2z_1^2(1-z_1)^2} \hspace{-0.5 cm} \frac{\de \kt^2}{\left(\kt^2+z_1^2 m_1^2\right) \kt^{2\epsilon}}  \nonumber\\ &  \times  P^{(d)}_{gh} (z_1,\kt^2)
\delta\left(z_1 -(1-x) \right).
\end{align}
The bounds on the transverse momentum integral come from the fact that $\kt^2$ has an upper limit. This is determined by observing that in the quasi-collinear limit: 
\begin{align}
\label{eq: scalar product}
   2 p_1 \cdot k&= \frac{\kt^2+z^2 m_1^2}{z(1-z)}= 2\left(E_1E_{\vec k} -\vec{p}_1\cdot \vec{k}\right)
    \nonumber \\
   &\simeq\frac{z(1-z)q^2}{4}\left[\frac{4m_1^2}{(1-z)^2q^2}+\theta^2\right],
\end{align}
where $\theta$ is the splitting angle, from which it follows
\begin{equation}
\label{eq: bound}
\kt^2< q^2 z^2(1-z)^2 \theta_{\text{max}}^2.
\end{equation}
The choice of $\theta_{\text{max}}$ is immaterial for the accuracy of interest, thus we can safely set $\theta_{\text{max}}=1$. 
We can easily compute Eq. $\eqref{eq:function-j}$ 
\begin{align}
   &\frac{1}{\sigma^{(d)}_0}\frac{\de \sigma^{(C)}}{\de x}= \frac{\as \cf \Sep }{2\pi} \nonumber \\
   &\left[\frac{1+x^2}{(1-x)^{1+2\epsilon}}\left(-\ln{\xi_1} + \epsilon \frac{\pi^2}{6}+\frac{\epsilon}{2}\ln^2{\xi_1}\right) 
    -2\frac{(1-\epsilon  \ln \xi_1)}{(1-x)^{1+2\epsilon}}
    \right].
\end{align}
This result can be rearranged by exploiting the identity
\begin{align}\label{eq: identity plus delta}
    &\frac{1}{(1-x)^{1+2\epsilon}}=-\frac{1}{2\epsilon}\delta(1-x) + \frac{1}{(1-x)_+} +\ord{\epsilon},
\end{align}
so that the final expression for the collinear function is
\begin{align}
    &\frac{1}{\sigma^{(d)}_0}\frac{\de \sigma^{(C)}}{\de x} = \frac{\as \cf S_\epsilon}{2\pi} 
\nonumber   \\
   &\left[ \delta(1-x)\left(\frac{\log{\xi_1} }{\epsilon} +\frac{1}{\epsilon} - \log{\xi_1} -\frac{1}{2}\log^2{\xi_1}  -\frac{\pi^2}{6}   \right) \right. \nonumber
   \\
   &\left. - \frac{1+x^2}{(1-x)_+} \log{\xi_1}-\frac{2 }{(1-x)_+} 
   \right].
\end{align}

Next, we address the computation of the anti-collinear function in the quasi-collinear limit. We have
\begin{align}
 &\label{eq:function-jbar}
\frac{1}{\sigma^{(d)}_0}\frac{\de \sigma^{(\bar C)}}{\de x} =\frac{\as}{2\pi} \frac{(4\pi \mur^2)^\epsilon}{\Gamma(1-\epsilon)}\int_0^1 \de z_2\int_{0}^{q^2 z_2^2(1-z_2)^2} \hspace{-0.5 cm}\frac{\de \kt^2}{\left(\kt^2+z_2^2 m_2^2\right) \kt^{2\epsilon}} \nonumber \\ &\times    P^{(d)}_{gh} (z_2,
\kt^2)
\delta\left( \frac{\kt^2+z_2^2 m_2^2}{q^2 z_2(1-z_2)} -(1-x)\right).
\end{align}
In this case using the $\delta$ function to compute the $\kt^2$ integral mixes both integration variables, thus rendering their evaluation non-trivial. It follows a restriction on the $z_2$ integral as now $1-x<z_2<1-a$ where 
\begin{equation}
    a= \frac{\xi_2}{1-x+\xi_2}.
\end{equation}
Again, using the identity in Eq. \eqref{eq: identity plus delta} we find
\begin{align}\label{eq: bar j final}
&\frac{1}{\sigma^{(d)}_0}\frac{\de \sigma^{(\bar C)}}{\de x} = \frac{\as \cf S_\epsilon}{2\pi} \nonumber
    \\
    &\left[ \delta(1-x)\left(\frac{\log{\xi_2} }{\epsilon} +\frac{1}{\epsilon} + \log{\xi_2} +\frac{1}{2}\log^2{\xi_2} +2 -\frac{\pi^2}{6}   \right) \right.
   \nonumber \\
&\left.-\frac{3}{2}\frac{1}{(1-x)_+} - \frac{2\log{(1-x+\xi_2)}}{(1-x)_+}    \right. \nonumber
\\ &\left. -\frac{1}{2(1-x)_+}\frac{\xi_2(2(1-x)+\xi_2)}{(1-x+\xi_2)^2}
\right].
\end{align}
We note that in this expression the $x\to 1$ and $\xi_2\to 0$ limits do not commute. For instance, focusing on the logarithmic term, we observe that by taking the soft limit first, we get a logarithmically enhanced but finite term, whereas this is divergent when considering the limits in the reverse order.
Similarly, also the last term is enhanced but finite as the limit $x \to 1$ is taken first, while it vanishes as $\xi_2 \to 0$. This interplay between the soft and massless limits is exactly the behaviour discussed at length in \cite{Corcella:2004xv, Gaggero:2022hmv, Ghira:2023bxr}.

Finally, the subtracted soft contribution vanishes at this order, as detailed in \ref{app: soft sub}.
Summing all contributions together all the infrared poles cancel and we obtain that, in the quasi-collinear limit, the $\ord{\as}$ contribution to the differential cross-section reads
\begin{align}
	\label{eq: total result massive}
	&\frac{1}{\sigma_0}\frac{\de \sigma^{(CV\bar C)}}{\de x} = 
	\frac{\as \cf}{2\pi} 
    \nonumber \\
    &\left[ \delta(1-x)\left(\pi^2-2 + \log^2\xi_2 +\frac{1}{2}\log\xi_2\right) \right.
    \nonumber \\
    &\left.
    -\log\xi_1 \left(\frac{1+x^2}{(1-x)_+}+\frac{3}{2}\delta(1-x)\right) -\frac{7}{2}\frac{1}{(1-x)_+} \right.
    \nonumber \\
	&\left.  - \frac{2\log(1-x+\xi_2)}{(1-x)_+} -\frac{1}{2(1-x)_+}\frac{\xi_2(2(1-x)+\xi_2)}{(1-x+\xi_2)^2} + \dots  \phantom{\frac{A^a}{A^a}} \hspace{-0.45 cm}
	\right].
\end{align}
Let us make some comments. 
First, we note that the coefficient of $\log\xi_1$ is, as expected, the DGLAP splitting function. Second, we observe that if we take the limit $1-x\ll \xi_2$, we recover Eq. \eqref{eq: full fo}, provided that we set $\xi_1=\xi_2=\xi$. 
Lastly, we notice that at the level of the total differential cross-section, the $\xi_2\to 0$ limit remains ill-defined. Nonetheless, from a physical point of view, this limit has to be smooth. We expect a divergent behaviour only for $\xi_1\to 0$ since we are examining the energy spectrum of the $h$-quark. We can solve this apparent contradiction by exploiting the following identity between plus distributions
 \begin{align}
     \left(\frac{1}{1-x}\right)_+ g(x)&= \left(\frac{g(x)}{1-x}\right)_+ \nonumber \\ &+\delta(1-x) \int^1_0 \de x'\,  \left(\frac{1}{1-x^\prime}\right)_+ g(x') ,
 \end{align}
from which we derive the following relations for the cases of interest:
\begin{align}
    \frac{\log(1-x+\xi_2)}{(1-x)_+} &= \left(\frac{\log(1-x+\xi_2)}{1-x}  \right)_+ 
    \\
    &+\delta(1-x)\left(\frac{1}{2}\log^2\xi_2  +\frac{\pi^2}{6}+\ord {\xi_2}\right), \nonumber
 \end{align}
 and
 \begin{align} 
\frac{1}{(1-x)_+}\frac{\xi_2(2(1-x)+\xi_2)}{(1-x+\xi_2)^2}& =\delta(1-x) \left(1+\log\xi_2 +\ord {\xi_2}\right) 
\nonumber \\ 
&+\left( \frac{1}{(1-x)}\frac{\xi_2(2(1-x)+\xi_2)}{(1-x+\xi_2)^2}  \right)_+,
\end{align}
Thus, we can rewrite the differential cross-section as:
\begin{align}
\label{eq: final expression x space}
    &\frac{1}{\sigma_0} \frac{\de \sigma^{(CV\bar C)}}{\de x}= \frac{\as \cf }{2\pi} 
    \nonumber \\
     &\left[ \delta(1-x)\left(\frac{2\pi^2}{3} -\frac{5}{2}\right) - \log \xi_1 \left(\frac{1+x^2}{(1-x)_+}
     +\frac{3}{2}\delta(1-x)\right) \right.
     \nonumber \\
    & \left.-\frac{7}{2}\frac{1}{(1-x)_+}
    -2\left( \frac{\log (1-x+\xi_2)}{1-x} \right)_+ 
    \right. \nonumber \\
    &\left.
  \quad \quad \quad \quad -\frac{\xi_2}{2}\left(  \frac{2(1-x)+\xi_2}{(1-x+\xi_2)^2} \frac{1}{(1-x)}  \right)_+ + \dots \right],
\end{align}
from which it is easy to see that the $\xi_2 \to 0$ limit is now well-defined. Furthermore, when $\xi_2 \to 0$, we fully recover the result in the massless scheme, see Eq. \eqref{eq: CC-fixed order}. More specifically, we are able to recover not only the logarithmic structure but also the coefficient of the delta function of the massless-scheme calculation in Eq. \eqref{eq: CC-fixed order}. This is the result we have been aiming for.

\section{$\NLLp$ resummation} \label{sec: numerics}
In this section, we exploit the results obtained above to extend our Mellin-space resummation formula to NLL$^\prime$. 
This is achieved by including in the resummed expression $\ord{\as}$ contributions that are not enhanced at large $N$ and that reproduce the correct constant behaviour of the massless and massive schemes.

\subsection{Mellin transform}
We start by analyzing Mellin moments of Eq. (\ref{eq: final expression x space}). The Mellin transform of the term $1/(1-x)_+$ is straightforward to compute and the explicit result is
\begin{align}
\int^1_0 \de x ~x^{N-1} \frac{1}{(1-x)_+}= -\log{\bar N} +\ord{\frac{1}{N}}.
\end{align}
The Mellin transform of the remaining plus distributions is far more difficult to compute as they depend also on the scale $\xi_2$. However, since we are interested in the large $N$ limit, we can employ the formal identity \cite{Catani:1990eg,Catani:1989ne, Catani:2003zt}
\begin{align}
\label{eq: approx Mellin}
    x^{N-1}-1\to -\widetilde{\Gamma}\left(1-\frac{\partial}{\partial \log \bar N}\right)\Theta\left(1-x-\frac{1}{\bar N}\right) +\ord{\frac{1}{N}},
\end{align}
where the operator $\widetilde{\Gamma}$ is defined as
\begin{equation}
    \widetilde{\Gamma} \left(1-\frac{\partial}{\partial \log \bar N}\right)= \exp\left\{ \sum^\infty_{n=2} \frac{\zeta(n)}{n} \left(\frac{\partial}{\partial \log \bar N}\right)^n\right\}.
\end{equation}
Through Eq. (\ref{eq: approx Mellin}) we are able to catch not only the logarithmic structure but also the constant terms. We find
\begin{align} \label{eq: Mellin transf1}
&-\widetilde{\Gamma} \left(1-\frac{\partial}{\partial \log \bar N}\right) \int_0^{1-\frac{1}{\bar N}} \de x
\;\frac{\log(1-x+\xi_2)}{1-x}   \nonumber \\
&= \, \text{Li}_2\left(-\frac{1}{\xi_2}\right) - \text{Li}_2\left(-\frac{1}{\bar N \xi_2}\right) - \log\bar N\log \xi_2 
\nonumber \\ & 
+\frac{\pi^2}{12}\frac{1}{(1+\bar N \xi_2)}  + F_1(\bar N \xi_2),
\end{align}
and
\begin{align} \label{eq: Mellin transf2}
&-\widetilde{\Gamma} \left(1-\frac{\partial}{\partial \log \bar N}\right) \int_0^{1-\frac{1}{\bar N}} \de x\;
 \frac{2(1-x)+\xi_2}{(1-x+\xi_2)^2} \frac{\xi_2}{(1-x)} 
\nonumber \\
& = -\frac{\bar N \xi_2 }{1+\bar N \xi_2} -\log(1+\bar N\xi_2)+F_2(\bar N \xi_2).
\end{align}
We are not able to express $F_1$ and $F_2$ in a closed form. However, we note that in the large-$N$ limit, they both vanish as an inverse power of $N$. Furthermore, they also both vanish in the $\xi_2 \to 0$ limit.

We can use the above results to build a $\NLLp$ resummed expression, which has the same form as Eq. \eqref{eq:FF-sep modified NLL}, but with a coefficient function that reproduces the correct result in both massive and massless calculations. Therefore, setting $\xi_1=\xi_2=\xi=m^2/q^2$, we write
\begin{align}\label{eq:FF-sep modified}
		&\widetilde{\sigma}^{(\text{NLL}^\prime)}(N, \xi)
        =  \widetilde{\mathcal{E}}^{(\text{sub})}(N,m^2,q^2) 
         \nonumber\\ &
				 \left(1+ \frac{\as(\mu^2) }{\pi}\left(\mathcal{C}_0^{(1)}+ \delta \mathcal{C}_0^{(1)} \right)\right) \left(1+ \frac{\as(\muOr^2) }{\pi}\mathcal{D}_0^{(1)} \right) 
                   \nonumber\\ &
				 \Exp \left[j\left(N,\xi,\frac{\mur^2}{q^2},\frac{\muOr^2}{m^2}\right)+\bar{j}\left(N,\xi,\frac{\mur^2}{q^2},\frac{\muOr^2}{m^2}\right)\ \right],
\end{align}
where
\begin{align}\label{eq: delta C1}
        &\delta \mathcal{C}_0^{(1)} = \cf\left[ \left(\frac{\pi^2}{12} + \frac{1}{4}\right) \frac{y}{1+y}
  \right. \nonumber\\
   &\left. +\Theta\left(1-y \right) \,\left(-\text{Li}_2\left(-y\right) + \frac{1}{4} \log(1+y)  \right) \right.\nonumber \\
  &\left.+\Theta\left(y-1 \right)\left(\frac{\pi^2}{6}+\text{Li}_2\left(-\frac{1}{y} \right)  + \frac{1}{4} \log\left(\frac{1+y}{y}\right)  \right) \right],
\end{align}
with $y= \bar N \xi$.
Eq. ~\eqref{eq: delta C1} is obtained from Eqs.~\eqref{eq: Mellin transf1} and~\eqref{eq: Mellin transf2} subtracting the logarithmic terms already present in the resummed exponent. Furthermore, we have also neglected the contributions arising from $F_1$ and $F_2$, as they do not affect the asymptotic behaviour of the coefficient function.  
Finally, we note that the value of the scale $\mu$ is not fixed by our NLL$^\prime$ calculation.

We note that Eq.~\eqref{eq:FF-sep modified} coincides with the resummed expression derived in~\cite{Cacciari:2001cw} when the limit $\xi\to 0$ is taken, because in this limit $\delta \mathcal{C}_0^{(1)} \to 0$, $j\to J$ and $\bar j \to \bar J$. Conversely, when $N\to \infty$, $\widetilde{\mathcal{E}}^{(\text{sub})}$ does not contribute, and, because
\begin{align}
    \delta \mathcal{C}_0^{(1)} \xrightarrow[N \to \infty]{} \frac{\cf}{4}(\pi^2 +1),
\end{align}
we correctly recover Eq.~\eqref{eq: total result massive}, including the $\ord{\as}$ constant 
\begin{equation}
\mathcal{C}_0^{(1)}+   \delta \mathcal{C}_0^{(1)}+\mathcal{D}_0^{(1)}\xrightarrow[N \to \infty]{} \cf \left(\frac{\pi^2}{2}-1 \right).
\end{equation} 

\begin{figure}
    \centering
    \includegraphics[width=1.03\linewidth]{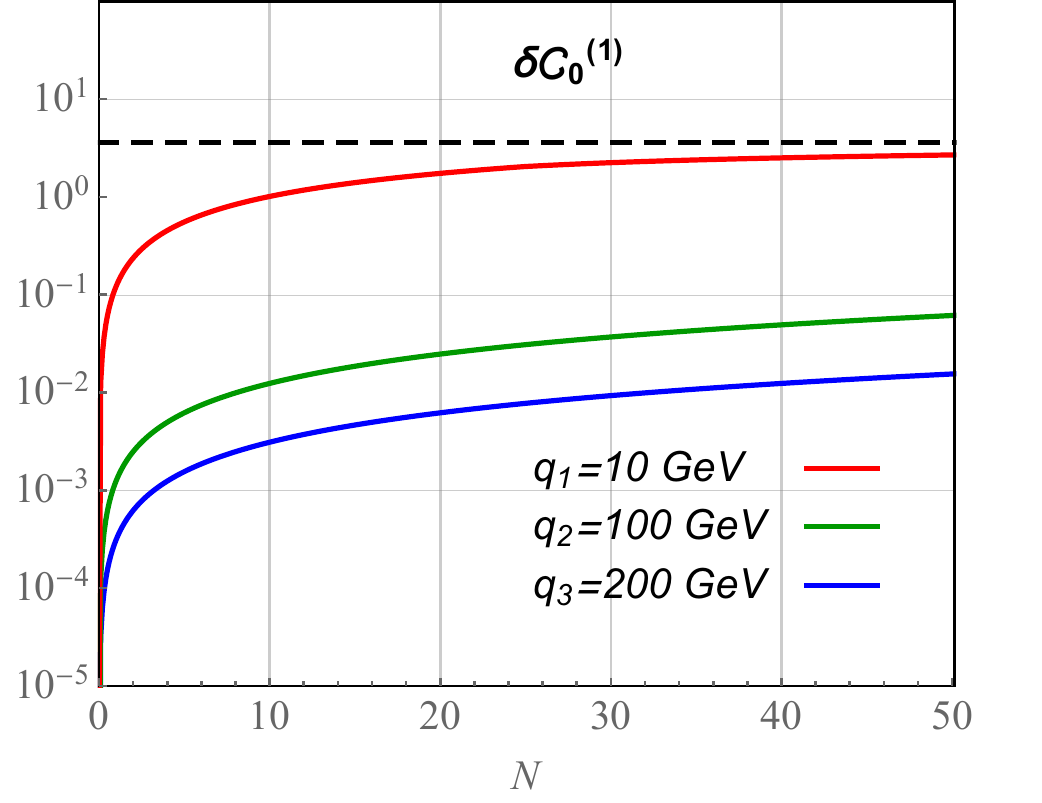}
    \caption{Correction factor $\delta \mathcal{C}_0^{(1)}$ in Mellin space for the three different energy values $q_1=10$ GeV (solid red),  $q_2=100$ GeV (solid green) and $q_3=200$ GeV (solid blue). The black dashed line corresponds to the asymptotic $N \to \infty$ value $\frac{\cf}{4}(\pi^2+1)$.}
    \label{fig: deltaC}
\end{figure}
The behaviour of $\delta\mathcal{C}_0^{(1)}$ is displayed in Fig.~\ref{fig: deltaC} for three different energy values $q_1=10$ GeV,  $q_2=100$ GeV and $q_3=200$ GeV. We consider the charm quark as the heavy quark with $m_c= 1.5$ GeV. As expected from Eq.~\eqref{eq: delta C1}, the higher centre-of-mass energy, the slower the correction $\delta\mathcal{C}_0^{(1)}$ approaches its asymptotic value.

\subsection{Numerical results for the charm ratio}
In this section, we investigate the numerical impact of the contribution added to the resummation. 
As an example, we choose a fragmentation observable that has been extensively studied in the literature, namely the charm ratio. This is defined by considering Mellin moments of the $D$ meson energy spectrum at different centre-of-mass energies. In this ratio, non-perturbative contributions arising from the $c \to D$ hadronization process should cancel and we are left with a perturbative observable that we can describe as
\begin{align}
    R\left(N, q_\text{A},q_\text{B}\right)= \frac{\widetilde{\sigma}(N, q_\text{A})}{\widetilde{\sigma}(N, q_\text{B})}.
\end{align}
The scale $q_A$ is typically set at the $Z$ pole ($q_A=91. 2$ GeV), while $q_B$ at the $\Upsilon(4S)$ resonance ($q_B=10.6$ GeV).
This observable has been studied in \cite{Cacciari:2005uk,Bonino:2023vuz} in the massless scheme and, more recently, in \cite{Cacciari:2024kaa} with the inclusion of the mass effects in the resummation of the coefficient function at NLL. 

In our current analysis, we compare the plain NLL resummation of mass and soft logarithms against the NLL$^\prime$ improvement. 
Scales are set to their natural values and renormalization scales are varied up and down by a factor of 2, to estimate theory uncertainties. 
As already mentioned, the choice of the scale $\mu$ is beyond our accuracy. A possible choice would be to take $\mu^2\equiv \mur^2\simeq q^2$ since this is the natural scale for the coefficient function. 
On the other hand, we can partially include NNLL correction by adopting a dynamical scale choice. We can do this by noting that $\delta \mathcal{C}_0^{(1)}$ arises from the anti-collinear region in which the emission angle scales as $1/N$ but saturates at the value set by the dead-cone effect. This suggests the scale choice:
\begin{align}\label{eq: mu2 choice}
\mu^2 = \text{max}\left[\frac{q^2}{\bar N}, \, m_c^2\right],
\end{align}
which was also adopted in~\cite{Dhani:2024gtx}, in the context of transition corrections beyond NLL to jet substructure observables. 
With this choice, $\mu^2\approx q^2$ when $N$ is sufficiently small. In contrast,  $\mu^2$ saturates to $m_c^2$ for large values of $N$, which corresponds to the appropriate energy scale of the calculation in the massive scheme. We call this second choice \emph{approximate} NNLL (nNLL). 

We stress that this approximation should be taken with a grain of salt as we have no means to assess the size of NNLL corrections that we do not control. In this context, when performing scale variations to evaluate the theoretical uncertainty, we only add counterterms to preserve NLL accuracy. Therefore, we do not expect any significant reduction in the scale dependence when going from NLL to nNLL.

\begin{figure}
    \centering
    \includegraphics[width=0.95\linewidth]{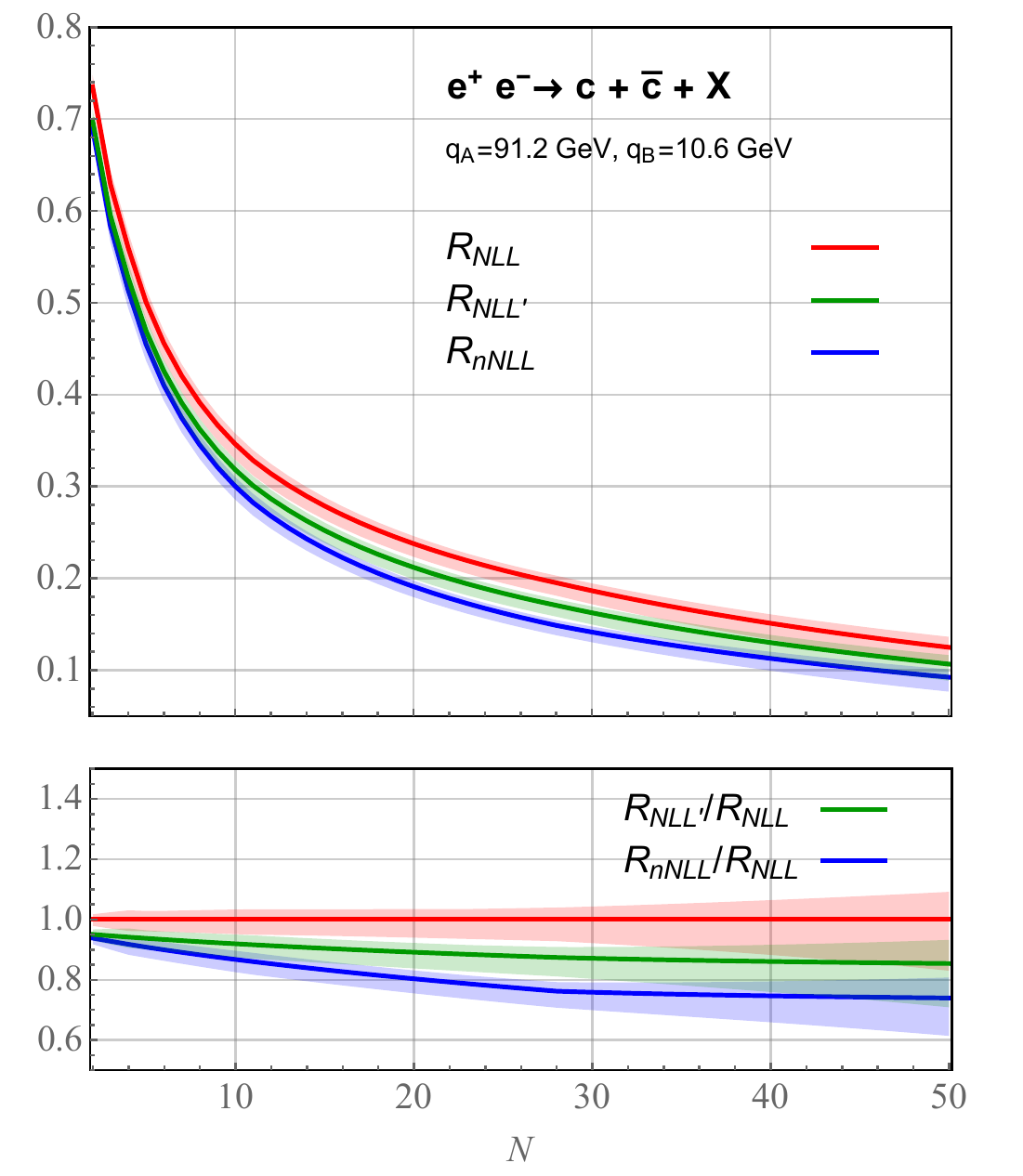}
    \caption{Comparison in Mellin space for the charm ratio $R(N)$. The first panel shows absolute predictions at NLL (solid red), NLL$^\prime$  (solid green) and nNLL  (solid blue). The second panel shows the relative corrections of the  NLL$^\prime$ and nNLL predictions with respect to NLL.}
    \label{fig:plot NLL NLLp nNLLp}
\end{figure}

In the main plot of Fig.~\ref{fig:plot NLL NLLp nNLLp}, we show three different curves for the charm ratio. The red curve corresponds to the prediction at NLL accuracy (computed with Eq.~\eqref{eq:FF-sep modified NLL}), while $R_{\text{NLL}^\prime}$ is its improvement at $\text{NLL}^\prime$.
Finally, $R_{\text{nNLL}}$ denote the approximate NNLL calculation. 
We note that the blue curve lies below the others as a consequence of the choice of the scale in Eq. (\ref{eq: mu2 choice}). 
In the sub-panel of the same plot we show the ratios of the $\NLLp$ (green) and nNLL (blue) predictions with respect to NLL.
We observe that the theoretical uncertainties associated to scale variations are not significantly affected by the inclusion of these additional corrections, as anticipated. The resummed theoretical predictions are modified by up $\sim 10 \%$ for $\NLLp$ and $\sim 20\%$ for nNLL already at $N=20$, while for $N=50$ these corrections respectively increase to $\sim 15\%$ and $\sim 25\%$. This seems to suggest that effects beyond NLL included in both  $\text{NLL}^\prime$ and nNLL are sizeable and the extension to a complete NNLL resummation of mass and soft logarithms can significantly enhance the accuracy on the theoretical side.

\section{Conclusions and Outlook}\label{sec: conclusion}

In this paper, we have considered soft-gluon resummation in heavy-flavour production in $e^+e^-$ annihilation. In particular, we have extended the formalism that allows one to consistently resum both mass and soft logarithms~\cite{Ghira:2023bxr} to NLL$^\prime$ accuracy.

We have achieved this result by computing the NLO differential cross section as the sum of virtual, soft, collinear and anti-collinear contributions, the latter ones evaluated in the quasi-collinear limit.  We have used dimensional regularization and the $\MSbar$ renormalization scheme. Thanks to a careful treatment of the plus-distributions, we have been able to construct an expression that smoothly reduces to known results, depending on the order of the subsequent $x\to1$ and $\xi \to 0$ limits. 
We have then considered Mellin moments of this result and used it to promote the accuracy of the resummation first derived in~\cite{Ghira:2023bxr} from NLL to NLL$^\prime$.

In order to assess the impact of these corrections, we have considered the charm-ratio observable. Here we have found that the resummed theoretical prediction is modified by $\sim$ 15 \% for large values of $N$, while the theoretical uncertainty as measured by scale variations, is not significantly affected.  
Additionally, we have estimated effects beyond NLL by including running coupling corrections. The impact on the charm ratio is sizeable, leading to an additional suppression at large $N$.  
This suggests that such contributions are not negligible and
because current theoretical estimates~\cite{Cacciari:2005uk,Bonino:2023icn,Cacciari:2024kaa} overshoot the experimental data, it strongly motivates the calculation of full NNLL corrections. Recent progress in the calculation of massive splitting kernels~\cite{Dhani:2023uxu,Craft:2023aew}, makes this task achievable in the near future.

\paragraph{Acknowledgments.}
We thank Prasanna Dhani, Giovanni Ridolfi, Gregory Soyez and Leonardo Vernazza for many useful discussions. 
This work is supported by the Italian Ministry of Research (MUR) under grant PRIN 2022SNA23K and by ICSC Spoke~2 under grant BOODINI. 
AG and SM also acknowledge support from the Munich Institute for \\
Astro-, Particle and BioPhysics (MIAPbP), which is funded by the Deutsche Forschungsgemeinschaft (DFG, German Research Foundation) under Germany's Excellence Strategy – EXC-2094 – 390783311.

\appendix

\section{NLL resummation explicit formulae} \label{app: nll explicit formulas}

In this appendix, we collect explicit expressions in Mellin space for the different objects that appear in the NLL resummation, Eq.~\eqref{eq:FF-sep}.

We separate the logarithmically enhanced contribution to the DGLAP evolution kernel $\widetilde{\mathcal{E}}$ from its regular component at large \( N \). 
To this end, we define a subtracted operator $\widetilde{\mathcal{E}}^{(\text{sub})}(N, \muOf^2, \muf^2)$  through
\begin{equation}
\label{eq:DGLAPsubtracted}
\widetilde{\mathcal{E}}(N,\muOf^2,\muf^2) = 
\widetilde{\mathcal{E}}^{(\text{sub})}(N,\muOf^2,\muf^2)  \,\Exp\left[E(N,\muOf^2,\muf^2)\right],
\end{equation}
where
\begin{equation}\label{eq:evolution operator integral}
E(N,\muOf^2,\muf^2)=-\int^{\muf^2}_{\muOf^2} \frac{\de k^2}{k^2}  \left\{A(\as(k^2)) \log \bar{N} + \frac{1}{2}B(\as(k^2)) \right\}.
\end{equation}
By construction, $\widetilde{\mathcal{E}}^{(\text{sub})}\to 1$ as $N\to\infty$.

The jet function $J$ defined in Eq.~(\ref{eq:FF-sep}) has the following form: 
\begin{equation}\label{eq:j-def}
J=D_0 + E+\Delta,
\end{equation}
with 
\begin{align}
\label{eq: delta function}
	\Delta\left(N,\frac{\muf^2}{q^2},\frac{\mur^2}{q^2}\right)=&\int_{\frac{1}{\bar{N}}}^1 \frac{\de z}{z}\int^{\muf^2}_{z^2q^2}\frac{\de k^2}{k^2} A(\as(k^2)),\\
\label{eq: initial condition}
D_0\left(N,\frac{\muOf^2}{m^2},\frac{\muOr^2}{m^2}\right)=& -\int_{\frac{1}{\bar{N}}}^1  \frac{\de z}{z} \Bigg\{ \int^{\muOf^2}_{z^2 m^2} \frac{\de k^2}{k^2} A(\as(k^2))\nonumber \\
+&H\left(\as\left(z^2 m^2\right)\right)\Bigg\}.
\end{align}
Finally, the massless jet function $\bar{J}$ reads:
\begin{align}
\label{eq: barJ}
	\bar{J}\left(N,q^2,\mur^2\right) &= \nonumber \\
    -&\int_{\frac{1}{\bar{N}}}^1  \frac{\de z}{z} \left\{\int^{zq^2}_{z^2q^2}\frac{\de k^2}{k^2} A(\as(k^2))+\frac{1}{2}B(\as\left(zq^2\right))\right\}.
\end{align}
Note that the resummed coefficient function that appears in Eq.~\eqref{fragm-master} is $C=\Delta+\bar J$.

The functions $A,B$ and $H$ have an expansion in powers of $\as$:
 \begin{equation}\label{Resummation_functions}
 	\begin{split}
 		A(\as)=&\sum_{k=1}^{\infty} \left(\frac{\as}{\pi}\right)^k A_{k},\quad	B(\as)=\sum_{k=1}^{\infty} \left(\frac{\as}{\pi}\right)^k B_{k},\\
 		H(\as)=&\sum_{k=1}^{\infty} \left(\frac{\as}{\pi}\right)^k H_{k},
 	\end{split}
 \end{equation}
 with
 \begin{equation}\label{Resummation_constants}
 	\begin{split}
 		A_{1}&=\cf,\quad A_{2}^{(n)}=\frac{\cf K^{(n)}}{2},\\
 		B_{1}&=-\frac{3}{2}\cf, \quad H_{1}=-\cf,
 	\end{split}
 \end{equation}
 and $K^{(n)}= \ca \left(\frac{67}{18}-\zeta_2\right)-\frac{5}{9} n$.
 The coefficients in Eq.~(\ref{Resummation_constants}) are sufficient to achieve NLL accuracy. Explicit expressions for the evolution factor $\mathcal{E}$, the coefficient function $\mathcal{C}$ and the initial condition for the fragmentation function $\mathcal{D}_0$ 
at the relevant level of accuracy are given for instance in \cite{Mele:1990yq, Cacciari:2001cw}.

\section{Soft contribution and double counting}\label{app: soft sub}

In this appendix, we report the main steps for the computation of the soft function, the expression of which can be directly determined by exploiting the Eikonal factorization:
\begin{align}\label{eq: soft function integral}
& \frac{1}{\sigma^{(d)}_0}\frac{\de \sigma^{(S)}}{\de x}= 8\pi \as \cf \mur^{2\epsilon}\int \frac{\de^{d-1}k}{(2\pi)^{d-1}2 E_{\vec{k}}} \delta \left(x- \frac{2 p_1\cdot q}{q^2}\right)\nonumber\\
&\left[\frac{p_1 \cdot p_2}{p_1 \cdot k \; p_2 \cdot k} -\frac{p_1^2}{2(p_1 \cdot k)^2}-\frac{p_2^2}{2(p_2 \cdot k)^2}\right].
\end{align}
Performing the integral in the small mass limit $\xi_{1,2} \ll 1$ we obtain
\begin{align}
\label{eq: soft function}
& \frac{1}{\sigma^{(d)}_0}\frac{\de \sigma^{(S)}}{\de x}=
\frac{\as \cf S_{\epsilon}}{2\pi}
\left[ \delta(1-x) \left(    \frac{  \log{\xi_1}}{\epsilon}  +\frac{  \log{\xi_2}}{\epsilon}  +  \frac{ 2}{\epsilon}  \right.\right.
\nonumber\\
\nonumber &\left.\left.
-\frac{\pi^2}{3}+2 -\frac{1}{2}\log^2{\xi_1}  +\frac{1}{2}\log^2{\xi_2}  - \log{\xi_1} + \log{\xi_2} \right) \right.\\
&\left. -\frac{2}{(1-x)_+}\left(\log{\xi_1} +\log{\xi_2} +2\right)
\right] .
\end{align}

The double counting which originates from the overlap of  soft and collinear sectors has to be subtracted from the total result. Such contribution can be computed by taking the soft limit of the collinear and anti-collinear functions in Eqs. \eqref{eq:function-j}--\eqref{eq:function-jbar}, which are
\begin{align}
\label{eq: j-soft}
	&\frac{1}{\sigma^{(d)}_0}\frac{\de \sigma^{(C)}}{\de x}\Big|_{\text{s}}= \frac{\as\cf}{2\pi} \Sep\nonumber \\
    &\left[\delta(1-x)\left( \frac{\log{\xi_1} }{\epsilon} +\frac{1}{\epsilon} - \log{\xi_1} -\frac{1}{2}\log^2{\xi_1}  -\frac{\pi^2}{6} \right)\right.
    \nonumber \\
        &\left.
	-\frac{2}{(1-x)_+} \left(\log\xi_1 +1 \right)    \right]
    \end{align}    
    \begin{align}
    \label{eq: bar j soft}
&\frac{1}{\sigma^{(d)}_0}\frac{\de \sigma^{(\bar C)}}{\de x}\Big|_{\text{s}}= \frac{\as\cf}{2\pi} \Sep
    \nonumber \\
    &\left[\delta(1-x)\left(\frac{\log{\xi_2} }{\epsilon} +\frac{1}{\epsilon} + \log{\xi_2} +\frac{1}{2}\log^2{\xi_2} +2 -\frac{\pi^2}{6} \right)\right.
    \nonumber \\
    &\left.
    -\frac{2}{(1-x)_+} \left(\log\xi_2 +1 \right)    \right].
\end{align}
It follows that 
\begin{equation}
   \frac{1}{\sigma^{(d)}_0}\frac{\de \sigma^{(S)}}{\de x}- \frac{1}{\sigma^{(d)}_0}\frac{\de \sigma^{(C)}}{\de x}\Big|_{\text{s}}-\frac{1}{\sigma^{(d)}_0}\frac{\de \sigma^{(\bar C)}}{\de x}\Big|_{\text{s}}=0.
\end{equation}
Therefore, Eq. \eqref{eq: total result massive} already provides the final result at $\ord{\as}$.

\addcontentsline{toc}{section}{References}

\bibliographystyle{jhep}
\bibliography{biblio}
\end{document}